\documentclass[12pt]{iopart}

\usepackage[dvipdfmx]{graphicx}
\usepackage{rotating}
\usepackage{amssymb}
\usepackage{mathptmx}
\usepackage[numbers]{natbib}
\usepackage {subfig}
\makeatletter
\bibpunct{}{}{,}{s}{}{,}

\begin{document}
%TC:ignore
%\newcommand{\hdblarrow}{H\makebox[0.9ex][l]{$\downdownarrows$}-}
\title{An optical transition-edge sensor with high energy resolution
}

\author{K. Hattori$^{a,b,c}$ \& T. Konno$^{a}$ \& Y. Miura$^{a}$ \& S. Takasu$^{a}$ \& D. Fukuda$^{a,c}$}

\address{$^{a}$ National Institute of Advanced Industrial Science and Technology,\\ Tsukuba, Ibaraki 305-8563, Japan\\% Tel.:\\ Fax:\\
$^{b}$ The International Center for Quantum-field Measurement Systems for Studies of the Universe and Particles (QUP), Tsukuba, Ibaraki 305-0801, Japan\\
$^{c}$ AIST-UTokyo Advanced Operando-Measurement Technology Open Innovation Laboratory, Tsukuba, Ibaraki 305-8563, Japan\\
\ead{kaori.hattori@aist.go.jp}}

\begin{abstract}
Optical transition-edge sensors have shown energy resolution %(typically 100 to 150 meV full width at half maximum, FWHM) 
for resolving the number of incident photons at the telecommunication wavelength. Higher energy resolution is required for biological imaging and microscope spectroscopy. In this paper, we report on a Au/Ti (10/20 nm) bilayer TES that showed high energy resolution. This was achieved by lowering the critical temperature $T_{\rm c}$ to 115 mK and the resultant energy resolution was 67 meV full width at half maximum (FWHM) at 0.8 eV. When $T_{\rm c}$ was lowered to 115 mK, the theoretical resolution would scaled up to 30 meV FWHM, considering that the typical energy resolution of optical TESs is 150 meV and $T_{\rm c}$ is 300 mK. To investigate the gap between the theoretical expectation (30 meV) and the measured value (67 meV), we measured its complex impedance and current noise. We found excess Johnson noise in the TES and an excess Johnson term $M$ was 1.5 at a bias point where the resistance was 10\% of normal resistance. For reference, the TES was compared with a TES showing typical energy resolution (156 meV FWHM). We will discuss what improved the energy resolution and what might have been the limiting factor on it. 
\end{abstract}
%\keywords{Transition-edge sensor, thermodynamic noise, complex impedance, single-photon detector, photon-number resolving detector}
\vspace{2pc}
\maketitle

%TC:endignore
\section{Introduction}

Transition-edge sensors can resolve the energy of a single photon by detecting a slight change in temperature caused by absorption of the photon~\cite{ref:TES0}. The detector design is optimized to detect photons of desired energy, ranging from gamma-ray, X-ray, UV, visible, and near-infrared~\cite{ref:tes-review}\cite{ref:optical_TES_1998}. We have been developing optical TESs capable of detecting visible and near-infrared photons. An optical TES can resolve the number of photons abosrbed in the detector simultaneously when a monochromatic light source is used. This photon-number resolution is important in fields such as quantum computing~\cite{ref:quantum-computing}, quantum information~\cite{ref:quantum-communication}~\cite{ref:TES-qcommunication}\cite{ref:TES-qcommunication2} and quantum metrology~\cite{ref:quantum-metrology}\cite{ref:quantum-metrology2}. In these fields, fast detector response is required. The size of optical TESs are set small to obtain sub-$\rm \mu$s responses. A TES has achieved the time constant of 150 ns~\cite{ref:tes-INRIM}. The small size also allows for efficient coupling to an optical fiber. 
The highest energy resolution was reported to be 105 meV ~\cite{ref:scanning_microscope2} full width at half maximum (FWHM), which is sufficient to identify the number of photons in the telecommunication wavelength bands.

Recently, new applications of optical TESs have been proposed. One such application is the combination of an optical TES and a scanning microscope~\cite{ref:niwa-frontier}~\cite{ref:scanning_microscope1}. In this system, two-dimensional color images were obtained by resolving the energy of each photon. When multi-color imaging is performed with photon detectors without the energy resolution, filters are generally required. 
To obtain color images with such the detectors, 
monochromatic images taken with filters transmitting light at wavelengths of interest are overlaid. The TES did not necessitate this procedure and provided multi-color images without filters.  

The TES had a low dark count rate and was able to be used for imaging under weak illumination. The TES was sensitive to photons at wavelengths ranging from near-infrared to visible light, and provided visible and near-infrared images in the wide bandwidth. This feature has the potential to open a new window for observation of biological samples. The new microscope system allowed the use of multiple fluorescent dyes for imaging~\cite{ref:scanning_microscope2}. The number of available dyes is determined by the energy resolution of a TES. The spacing between adjacent emission peak wavelengths of the dyes must be wider than the wavelength resolution.
This system can also be used for microscopic spectroscopy, where the microscope focuses on a fixed point and collects photons over a long period of time to obtain a spectrum. As the energy resolution of a detector becomes higher, a measured spectra can provide more information. 

In photon-number counting, requirement on the energy resolution is less stringent. A detector must exhibit the energy resolution high enough to avoid overlap between peaks of adjacent photon number states. To reduce an error in the photon number down to 1 \%, the peaks must be separated more than $5.2\sigma$, where $\sigma$ is the standard deviation of the peaks. For this purpose, the typical energy resolution of optical TESs (150 meV) is sufficient for use of 1550 nm telecommunication band. On the other hand, for biological imaging and microscopic spectroscopy, higher resolution is desired. 

The energy resolution $\Delta E_{\rm FWHM}$ is expressed in terms of the wavelength resolution $\Delta \lambda_{\rm FWHM} = \lambda^2\Delta E_{\rm FWHM}/hc$, where $\lambda$ is the wavelength of incident photons, $h$ is Planck constant and $c$ is the speed of light. When the energy resolution is 0.1 eV, the wavelength resolution at 550 nm (green) is 24 nm,
which may be comparable with spacing between adjacent emission peak wavelengths of dyes. To separate the peaks, the energy resolution must be improved.

The theoretical energy resolution of a TES is given by

\begin{equation}
    \Delta E_{\rm FWHM} = 2\sqrt{2\ln 2}
    \sqrt{4 k T_{\rm c}^2 \frac{C}{\alpha_I} \sqrt{n(1+2\beta _I)(1+M^2)/2}},
    \label{eq:delta-E-simple}
\end{equation}

where $k$ is the Boltzmann constant, $T_{\rm c}$ is the critical temperature, $C$ is the heat capacity, 
$\alpha _I = \frac{T}{R} \frac{\partial R}{\partial T}|_{I}$ is the temperature sensitivity,
$\beta _I = \frac{I}{R} \frac{\partial R}{\partial I}|_{T}$ is the current sensitivity,
$R$ is the resistance of the TES, $I$ is the current flowing through the TES, $n=5$ in the electron-phonon limited conductance, and $M$ expresses the excess Johnson noise term .
The energy resolution can be improved by lowering $T_{\rm c}$, the heat capacity $C$ and $\beta_I$, and increasing $\alpha_I$.
At present, there is no established method to control $\alpha_I$ and $\beta_I$.
To reduce the heat capacity, the size of a TES must be smaller. The size must be comparable or larger than the mode field diameter (MFD) of an optical fiber to obtain the high detection efficiency of photons. MFD of single-mode fibers is typically several $\rm \mu$m, ex., 3.2 $\rm \mu$m at 1550 nm for UHNA-7. When a TES is smaller than MFD, it enhances the coupling loss and deteriorates the detection efficiency~\cite{ref:TES-kobayashi}. The size of a TES is typically 8 ${\rm \mu m}\times8$ ${\rm \mu m}$, which is sufficient with MFD. 
To obtain high energy resolution with maintaining the detection efficiency, we chose to lower $T_{\rm c}$.
A low-$T_{\rm c}$ TES has a reduced heat capacity, which is proportional to $T_{\rm c}$ with a given volume.
From Eq.\ref{eq:delta-E-simple}, the energy resolution is proportional to $T^{1.5}_{\rm c}$.
The typical $T_{\rm c}$ and the energy resolution of our TESs are 300 mK and 150 meV, respectively. When this is lowered to 100 mK, the resolution is expected to be 30 meV.

In this paper, we will show that the energy resolution reached 67 meV FWHM at 0.8 eV by lowering $T_{\rm c}$ to 115 mK. The energy resolution was significantly enhanced but was did not reach the expected resolution of approximately 30 meV. To investigate the gap between the theoretical expectation and the measured value, the current noise of the detector was measured and compared with the theoretical noise. 
For TESs designed to detect X-ray photons, the excess Johnson noise is one of the limiting factors~\cite{ref:excess-noise}\cite{ref:excess-noise2}\cite{ref:excess-noise3}.
Here, we will show that there was the excess Johnson noise in the optical TES, and will discuss if it significantly contributed to deterioration of the energy resolution, and what were the major factors avoiding the energy resolution from reaching below 50 meV. 

We also tested an optical TES which showed typical energy resolution (156 meV FWHM at 1.46 eV), compared the two TESs to discuss their differences and similarities.

\section{Measurements and results}
\subsection{Optical transition-edge sensors}
In this study, we tested two TESs, one that achieved the high energy resolution (TES2 in Tab.\ref{tab:TES}) and other that showed typical resolution (TES1 in Tab.\ref{tab:TES}). They have Ti/Au thickness of 20/10 nm. Their size was 8 ${\rm \mu m}\times8$ ${\rm \mu m}$. The critical temperature ($T_{\rm c}$) of TES2 was 115 mK. $T_{\rm c}$ of TES1 was also fairly low, 143 mK. The TESs were cooled down in a dilution refrigerator. The bath temperature was set to 7 mK. The TESs were embedded in optical cavities optimized to maximize the detection efficiency at wavelengths of interest (950 nm for TES1 and 1550 nm for TES2). 
%As $T_{\rm c}$ gets lower, the detector response becomes slower, since the heat capacity $C$ is proportional to $T_{\rm c}$ and the thermal conductivity $G$ is proportional to $T_{\rm c}^{4}$, and the natural time constant $C/G$ is proportional to $T_{\rm c}^{-3}$. 

%A slow TES is suitable for complex impedance measurement. Important parameters such as $\alpha$ and $\beta$ can be extracted from measurements at low frequencies. For fast optical TESs, high-frequency measurements are necessary and coaxial cables should be installed to mitigate parasitic impedance. However, in this setup, a SQUID loop was set to open and the readout noise was fairly high. 

\begin{figure}[!t]
\begin{center}
\includegraphics[width=0.6\linewidth, keepaspectratio]{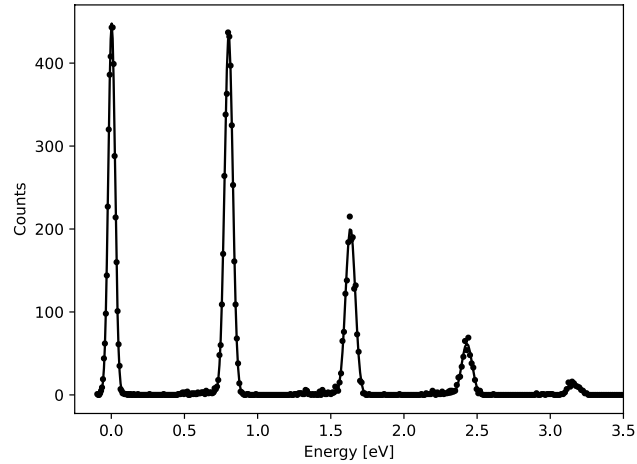}
\caption{TES2 response on a pulsed laser at 1550 nm (0.8 eV). The energy resolution of a Au/Ti bilayer TES (TES2 in Tab.~\ref{tab:TES}) reached 67 meV FWHM at 0.8 eV (the single-photon peak) by lowering Tc (115 mK).}
%DR-Run5-20200207/WF/unit3ch1
%python3 fit_peaks_LTD19.py 1550nm-60dB-1GHz-tes_histogram.dat
\label{fig-1550nm}
\end{center}
\end{figure}

The energy resolution was measured using pulsed lasers. Figure~\ref{fig-1550nm} showed response of TES2 to a pulsed laser at 1550 nm (0.8 eV). The energy resolution was 67 meV. Since TES1 was designed to detect photons at around 950 nm, an 850 nm (1.46 eV) pulsed laser was used. The measured energy resolution was 156 meV.

\begin{table*}[htb]
  \caption{Table of parameters regarding to TESs}
  \label{tab:TES}
  \centering
  \begin{tabular}{|c|c|c|} \hline
    & TES1 & TES2\\ \hline
    Material & Ti (20 nm) / Au (10 nm) & Ti (20 nm) / Au (10 nm)\\ \hline
    Size [$\rm{\mu}$m$^2$] & $8\times 8$ & $8\times 8$\\ \hline
    Critical temperature [mK] & 143 & 115\\ \hline
    Normal resistance [$\rm{\Omega}$]& 2.7 & 2.8 \\ \hline
    Energy resolution (FWHM) [meV] & 156 (at 1.46 eV) & 67 (at 0.8 eV)\\ \hline
    Thermal model & Single block & Two block \\ \hline
    $C/\alpha_I$ [J/K] at $0.1R_{\rm n}$ 
    & $4.1\times10^{-18}$ & $4.8\times10^{-19}$\\ \hline
    Excess Johnson noise term $M$ at $0.1R_{\rm n}$& 0.95 & 1.5 \\ \hline
    \end{tabular}
    \label{tab:TES}
    %python3 Z_TES3.py 1V.dat 0V.dat imp-100.dat 0.1 IV-unit3ch1-6p8mK_out.dat
    %python3 Z_TES3.py 1V.dat 0V.dat imp-100.dat 0.1 IV-unit4ch1-8mK_2_out.dat
\end{table*}

\subsection{Complex impedance}
\begin{figure*}[!t]
\centering
\subfloat[]{\includegraphics[width=0.25\linewidth, keepaspectratio]{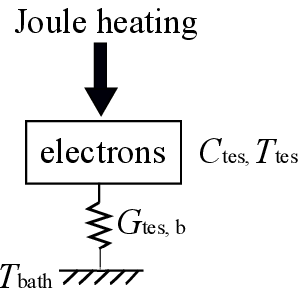}}
\hfil
\subfloat[]{\includegraphics[width=0.45\linewidth, keepaspectratio]{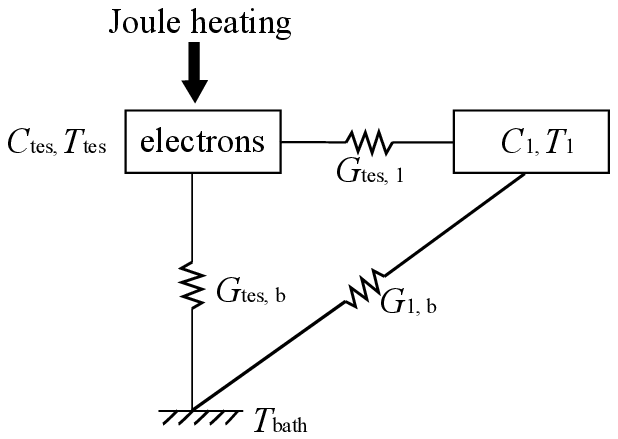}}
\hfil
\caption{ Schematic diagram of thermal models.
(a) A single-block model. (b) A two-block model.
%In both models, phonons in an optical TES are strongly coupled to a heat bath and can be effectively treated as the heat bath.
}
\label{fig:thermal-model}
\end{figure*}

We measured the complex impedance to extract parameters characterizing the TESs and to calculate the theoretical values for current noise and the energy resolution.
The complex impedance is determined by a thermal model of a TES.
The simplest model is a single-block model consisting of a TES and the thermal bath, as shown in Fig.~\ref{fig:thermal-model}(a).
This model described behavior of optical TESs well~\cite{ref:optical_TES_1MHz}~\cite{ref:Ztes-2020}.
%Phonons in a TES are strongly coupled with the thermal bath and their temperature is equal to the bath temperature.
%Therefore, phonons and the thermal bath can be treated as a thermal block.
The complex impedance of a single-block TES at an angular frequency $\omega$ can be written by~\cite{ref:TES0}, \cite{ref:Lindeman_Ztes}

\begin{equation}
  Z_{\rm TES}(\omega) = R(1+\beta _{\rm I})+\frac{R\cal{L}}{1-\cal{L}}
  \frac{2+\beta _{\rm I}}{1+i\omega \tau_{I{\rm ,s}}},
        \label{eq:Z_TES}
\end{equation}

where $\cal{L}$ is the constant-current loop gain given by ${\cal{L}} = R I^2 \alpha _I / G_{\rm tes,b}T$, $G_{\rm tes,b}$ is the thermal conductance between the TES and the thermal bath,
and $\tau_I$ is a time constant given by $\tau_{I{\rm ,s}} = \tau_0/(1-{\cal{L}})$,
where $\tau_0$ is an intrinsic time constant, $C/G_{\rm tes,b}$.

\begin{figure*}[!t]
\centering
\subfloat[]{\includegraphics[width=0.45\linewidth, keepaspectratio]{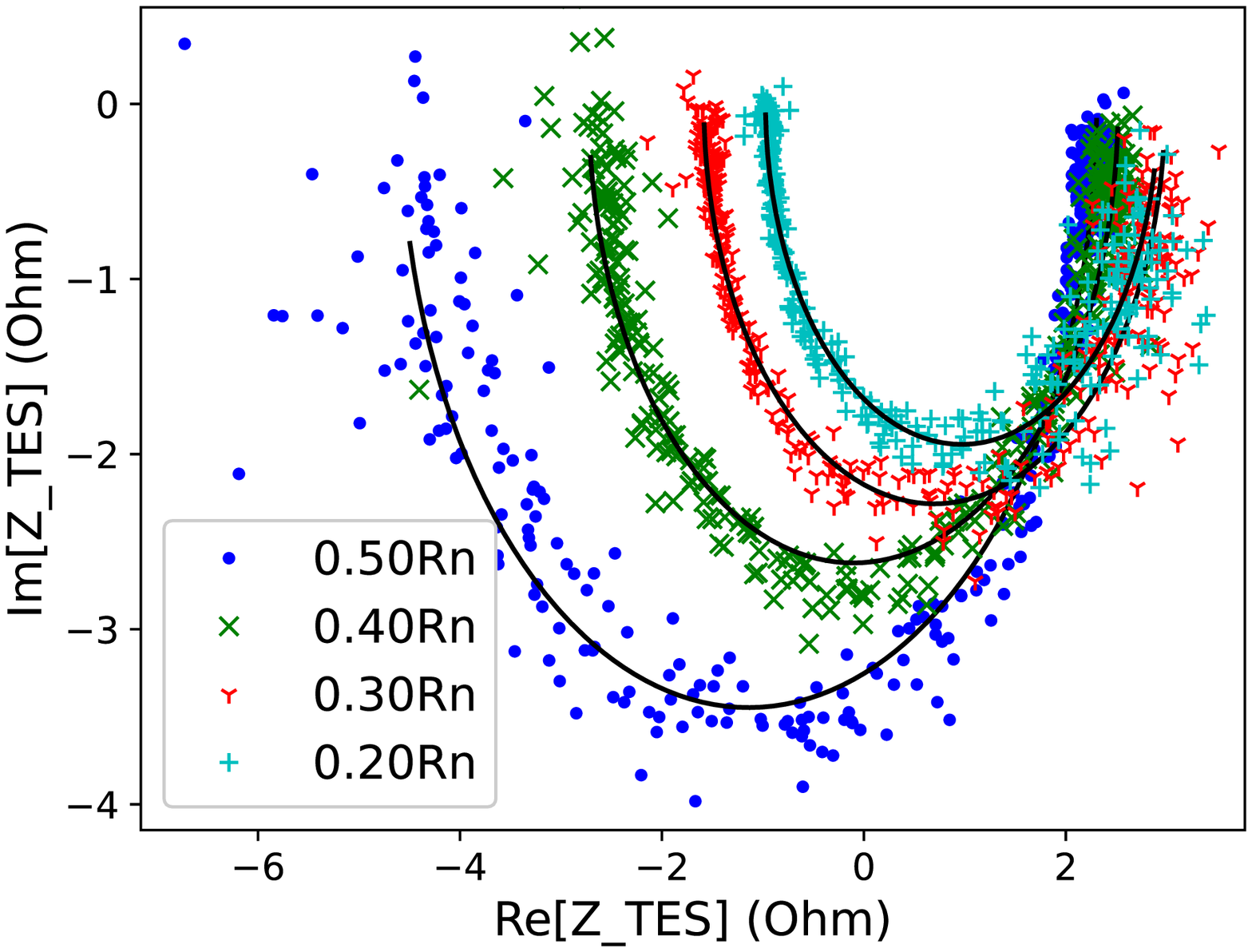}}
\hfil
\subfloat[]{\includegraphics[width=0.45\linewidth, keepaspectratio]{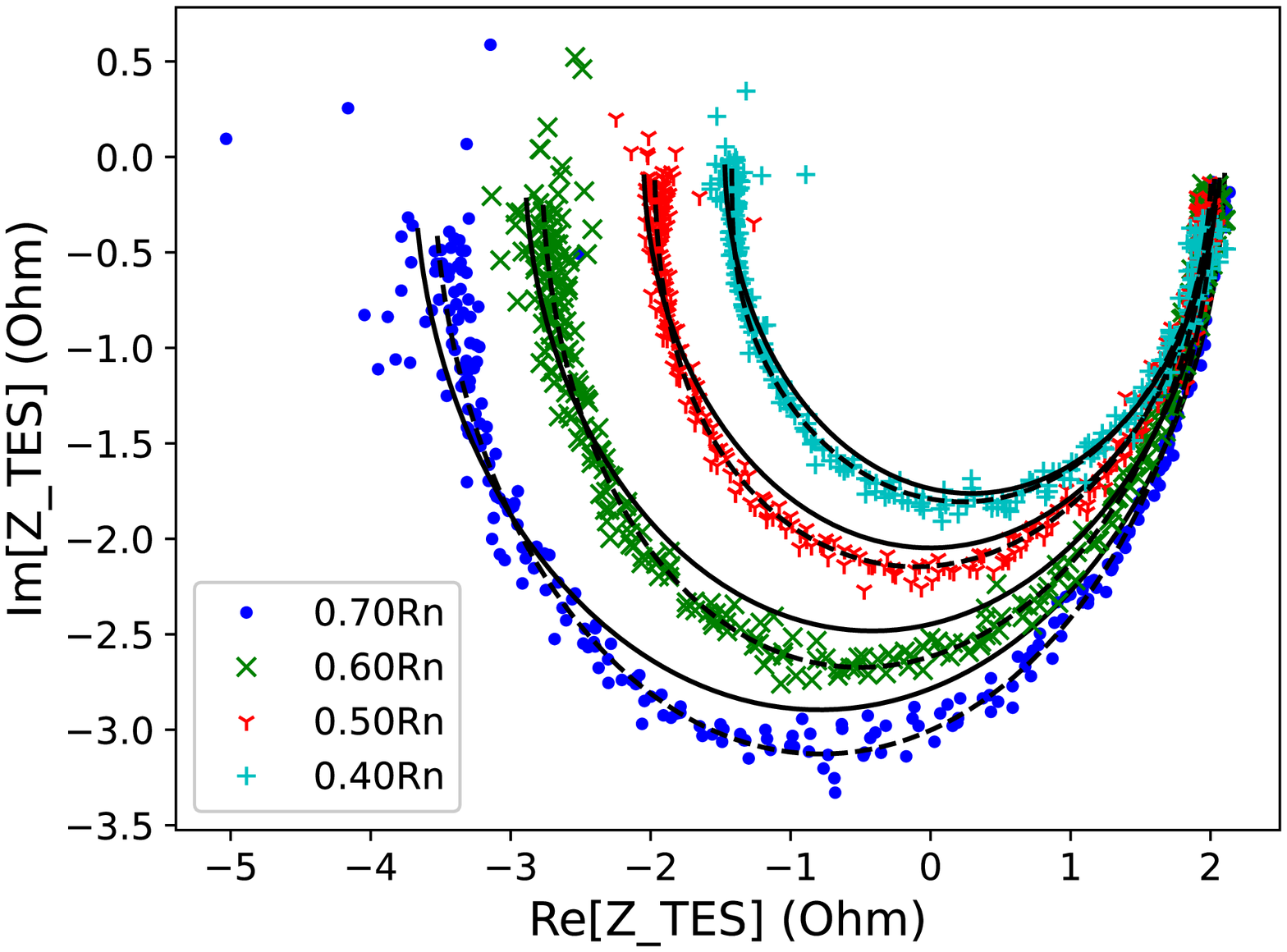}}
\hfil
\caption{ Measured complex impedance at various bias points.
(a) TES1 and (b) TES2.
The lines in (a) and (b) are fitted impedance from the single-block model (solid) and from the two-block model (dashed). $R_{\rm n}$ is the normal resistance.
}
%/home/tes/TES/data/DR-Run9-20200616/Imp/unit3ch1/6p8mK_20200701
%python3 ~/hattori/scripts3/fit_ZTES-plots.py imp-700_ZTES.dat imp-600_ZTES.dat imp-500_ZTES.dat imp-400_ZTES.dat
%DR-Run9-20200616/Imp/unit4ch1
%python3 fit_ZTES-plots_LTD19.py imp-500_ZTES.dat imp-400_ZTES.dat imp-300_ZTES.dat imp-200_ZTES.dat 
\label{fig:imp-TES}
\end{figure*}

\begin{figure*}[!t]
\centering
\subfloat[]{\includegraphics[width=0.45\linewidth, keepaspectratio]{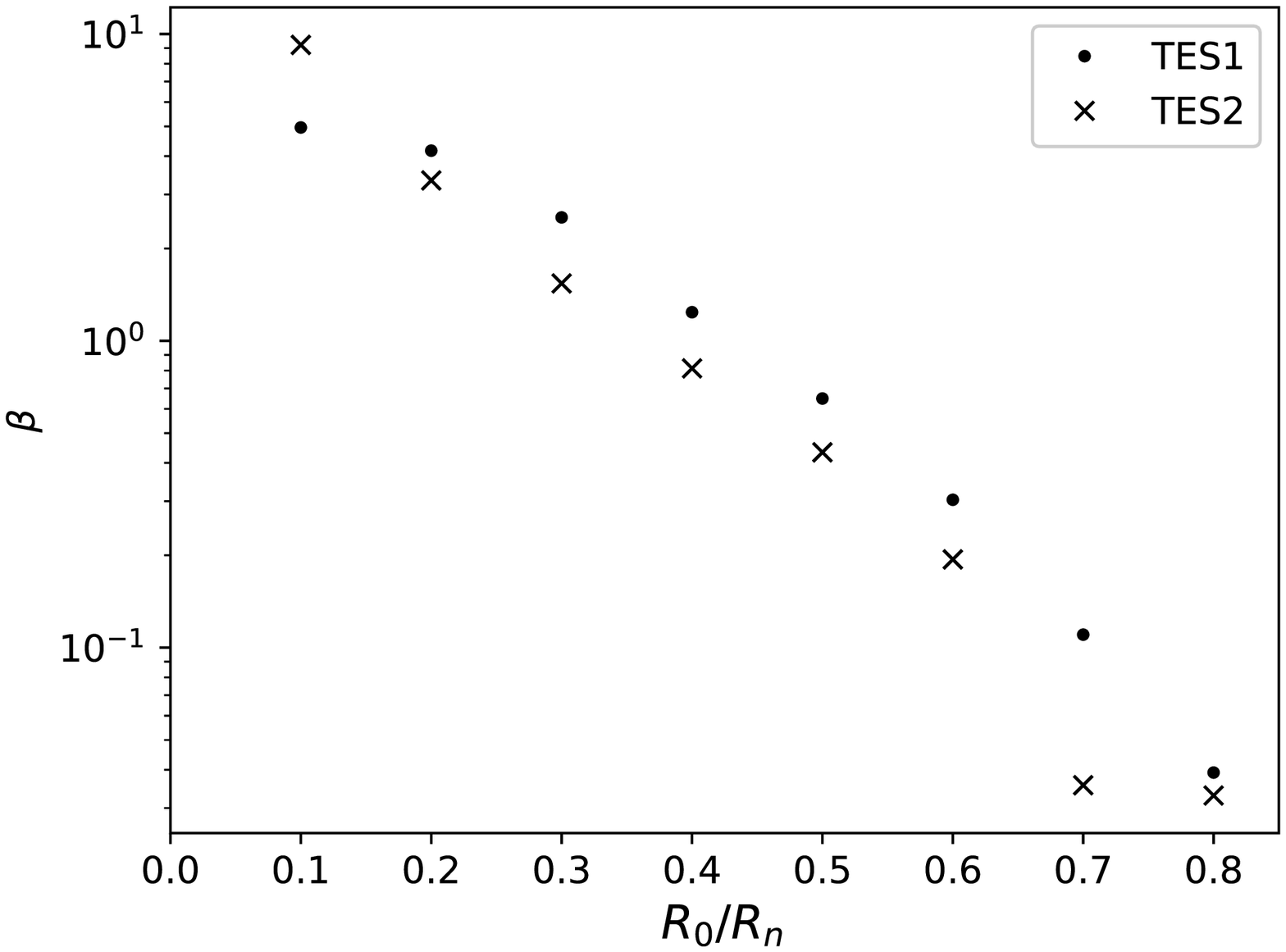}}
\hfil
\subfloat[]{\includegraphics[width=0.45\linewidth, keepaspectratio]{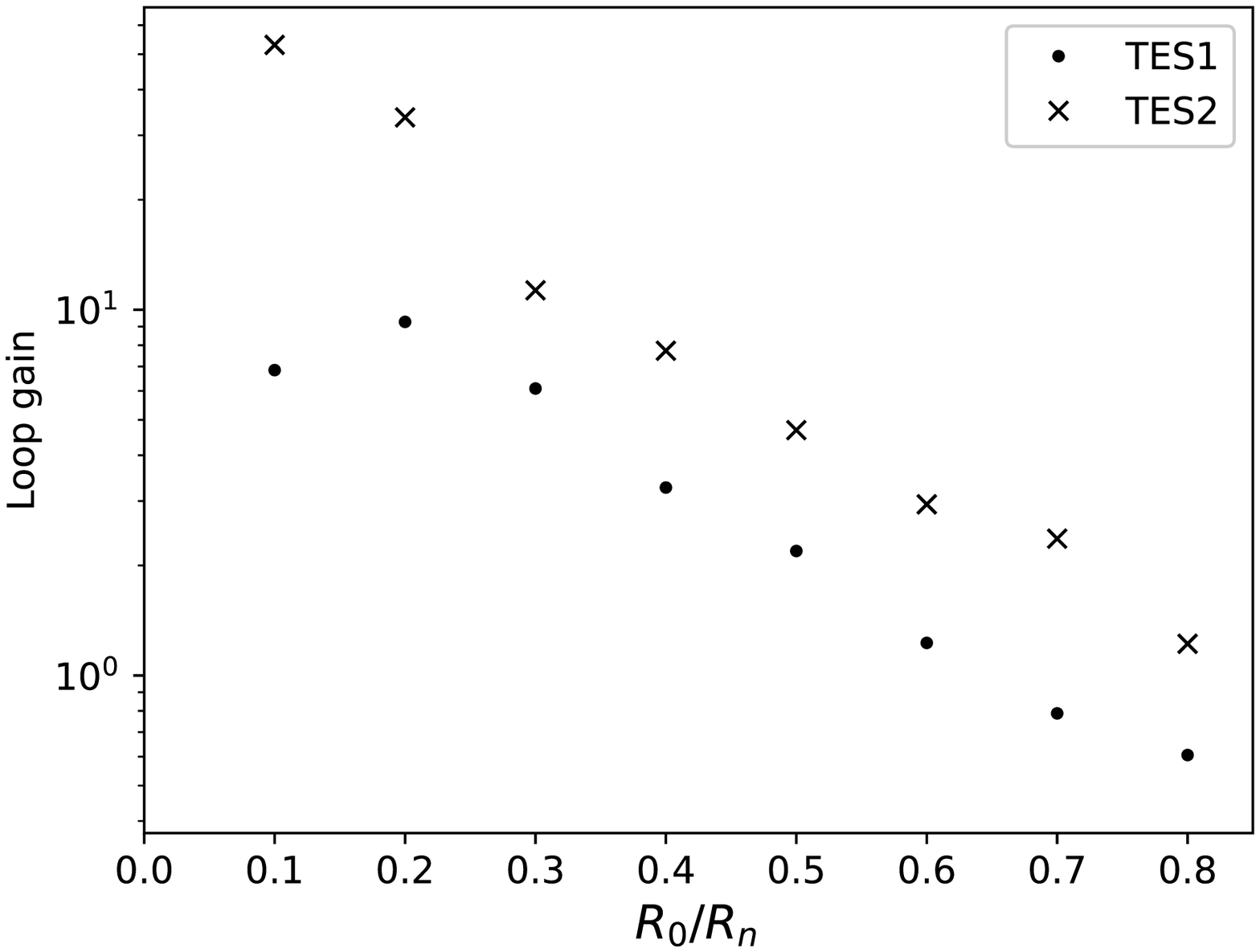}}
\hfil
\caption{Parameters extracted from measured complex impedance.
(a) Current sensitivity.
(b) Loop gain.
}
\label{fig:beta-loopgain}
%DR-Run9-20200616/WF
%python3 plot_data-loopgain-beta.py unit3ch1/6p7mK-20200701/results_20210823.dat 
%python3 plot_data-loopgain-beta.py unit4ch1/results_20200713.dat
\end{figure*}

The complex impedance was measured by injecting small signals from a network analyzer into a voltage-biased TES on a cold stage at 7 mK.
The current flowing through the TES was measured using a SQUID. 
%Twisted pair cables for signal transmission from the room temperature electronics to the cold readout and vice versa.
As shown in Fig.~\ref{fig:imp-TES}(a), the complex impedance of TES1 can be well described by Eq.~\ref{eq:Z_TES}. Figure~\ref{fig:beta-loopgain}(a) shows the extracted $\cal{L}$ and $\beta$. 

On the other hand, the responses of TES2 apparently deviated from that expected from the single-block model, as shown in Fig.~\ref{fig:imp-TES}(b)(solid lines). Instead, a two-block model (Fig.~\ref{fig:thermal-model}(b)) was adopted for TES2. The two-block model includes an additional thermal. When the thermal conductance between the additional body and the thermal bath is zero, the model represents a hanging model.This model describes a TES with an absorber~\cite{ref:excess-noise}~\cite{ref:Ztes-akamatsu}.  
When the thermal conductance between the TES and the thermal bath is zero, it represents an intermediate model~\cite{ref:Ztes_Maasilta}.
We will show that behavior of TES2 can be explained by the two-block model in Sec.\ref{sec:noise}. The complex impedance given by the model is written as~\cite{ref:Ztes_Maasilta}

\begin{equation}
  Z_{\rm TES}(\omega) = R(1+\beta _{\rm I})+\frac{R{\cal{L_{\rm eff}}}( 2+\beta _{\rm I})}{1-\cal{L_{\rm eff}}} /
  \left[ 1+i\omega \tau_I - \frac{\rm g_{two-body}}{(1-{\cal{L_{\rm eff}}})(1+i\omega \tau_1)}\right],
        \label{eq:Z_TES-2}
\end{equation}

where $\tau_1$ is the time constant of the additional body,  $\cal{L_{\rm eff}}$ is the effective loop gain and $g_{\rm  two-body}$ is a term as a function of thermal conductance in the system. These parameters can be written as

\begin{equation}
{\cal{L_{\rm eff}}} = RI^2\alpha_I/(G_{\rm tes,1}(T_{\rm tes}) + G_{\rm tes,b}),
\label{eq:L_eff}
\end{equation}

\begin{equation}
\tau_I = \frac{C_{\rm tes}}{(G_{\rm tes,1}(T_0) + G_{\rm tes,b})(1 - {\cal{L_{\rm eff}}})},
\label{eq:tau_I}
\end{equation}

\begin{equation}
\tau_1 = \frac{C_1}{G_{\rm tes,1}(T_1) + G_{\rm 1,b}},
\end{equation}

\begin{equation}
g_{\rm two-body} = \frac{G_{\rm tes, 1}(T_{\rm tes}) G_{\rm tes, 1}(T_{\rm 1})}  {(G_{\rm tes, 1}(T_{\rm tes}) + G_{\rm tes, b}) (G_{\rm tes, 1}(T_{\rm 1}) + G_{\rm 1, b})},
\end{equation}

where $C_1$ and $T_1$ are the heat capacity and the temperature of the additional body, respectively, $G_{\rm tes, 1}$ is the thermal conductance between the TES and the additional body, and $G_{\rm 1, b}$ is the thermal conductance between the additional body and the thermal bath.

As shown in Fig.~\ref{fig:imp-TES}(b)(dashed lines), the measured complex impedance of TES2 was fitted well with Eq.\ref{eq:Z_TES-2}. Assuming that $g_{\rm  two-body}$ and $\tau_1$ were independent of a bias point, they were 0.326 and 11.4 $\mu$s, respectively, as extracted from the measured complex impedance shown in Fig.~\ref{fig:imp-TES}(b). $\tau_1$ was similar to $\tau_0$ (13.9 $\mu$s at 0.5 $R_{\rm n}$). 
%All the two-block models have the same form of the complex impedance (Eq.\ref{eq:Z_TES-2}), but different forms of the current noise. To determine which of the three two-block models is appropriate for TES2, we measured the current noise. 

In TES2, while most of important parameters can be extracted from the complex impedance, some parameters such as the ratio between $G_{\rm tes,b}$ and $G_{\rm tes,1}$ remain unknown. Therefore, $G_{\rm tes,b}$ is unknown and $\alpha_I$ cannot be derived. However, $C/\alpha_I$, which is an important parameter for the energy resolution, can be directly extracted from the complex impedance. In the two-block model, $C/\alpha_I$ can be calculated using Eqs.~\ref{eq:L_eff}, \ref{eq:tau_I}, ${{\cal L}}_{\rm eff}$ and $\tau_I$. In the single-block model, from Eq.~\ref{eq:delta-E-simple}, the energy resolution is proportional to $\sqrt{C/\alpha_I}$, if assuming that the readout noise is zero. This is also the same applies to the two-block model, as shown in Appendix~\ref{app:E}. 
%Even if $\alpha_I$ is unknown, the theoretical value of the energy resolution can be calculated. 
The values of $C/\alpha_I$ extracted from the complex impedance measurements are shown in Tab.\ref{tab:TES}.
TES2 showed smaller $C/\alpha_I$ than TES1.
This is consistent with the fact that TES2 exhibited better energy resolution.
Using Eq.~\ref{eq:Z_TES-2}, the current sensitivity and the effective loop gain of TES2 were extracted from the complex impedance, as shown in Fig.~\ref{fig:beta-loopgain}(b). 
%The effective loop gain is defined by Eq.~\ref{eq:L_eff}. 
%To obtain further information associated with the thermal conductivity in TES2, 
%To determine a two-block model for TES2, the thermodynamic noise was measured. 
%However, the ratio between $G_{\rm tes,b}$ and $G_{\rm tes,1}$ still remains unknown, as shown in the following section. 

\begin{figure*}[!t]
\centering
\subfloat[]{\includegraphics[width=0.45\linewidth, keepaspectratio]{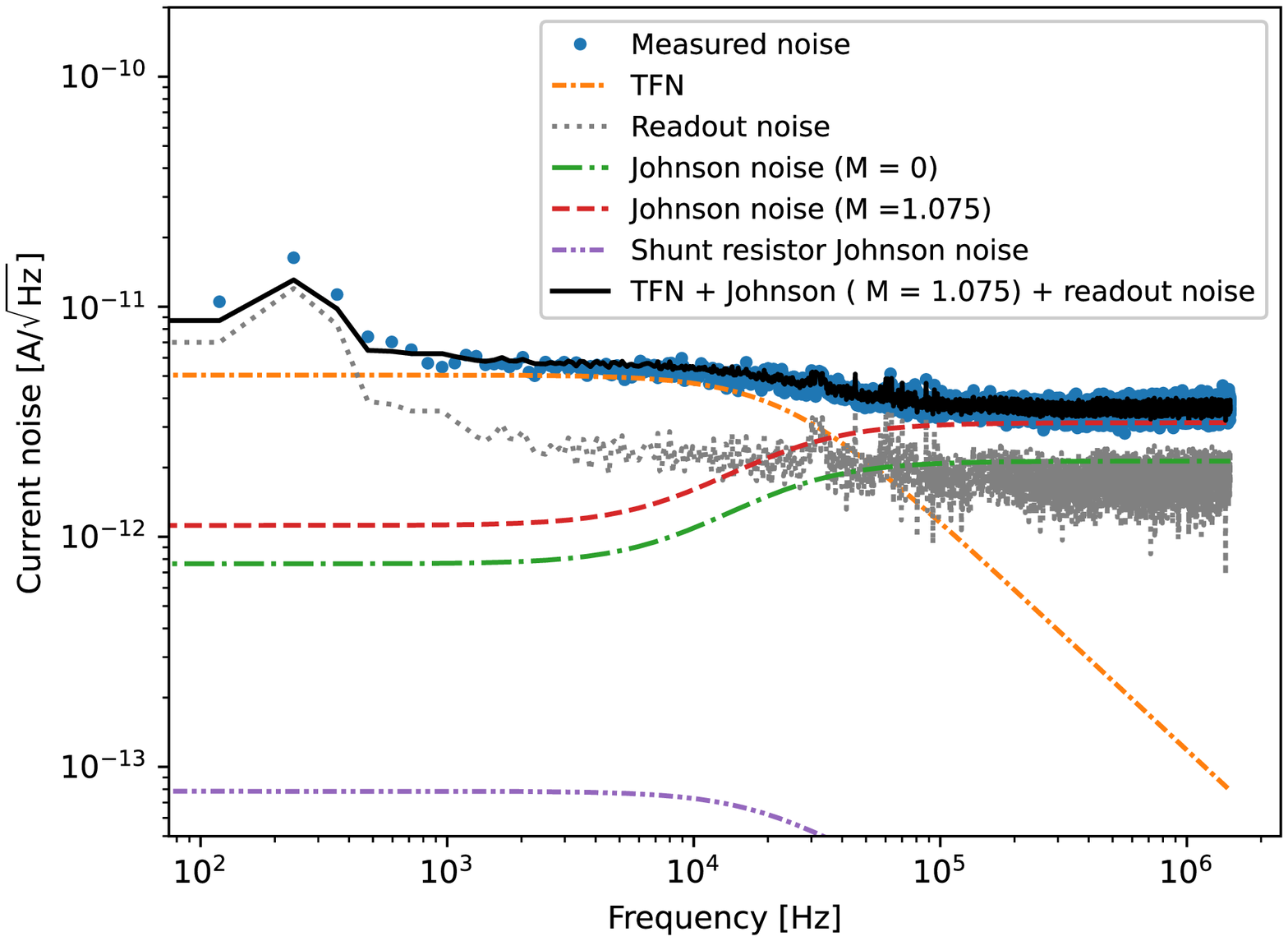}}
\hfil
\subfloat[]{\includegraphics[width=0.45\linewidth, keepaspectratio]{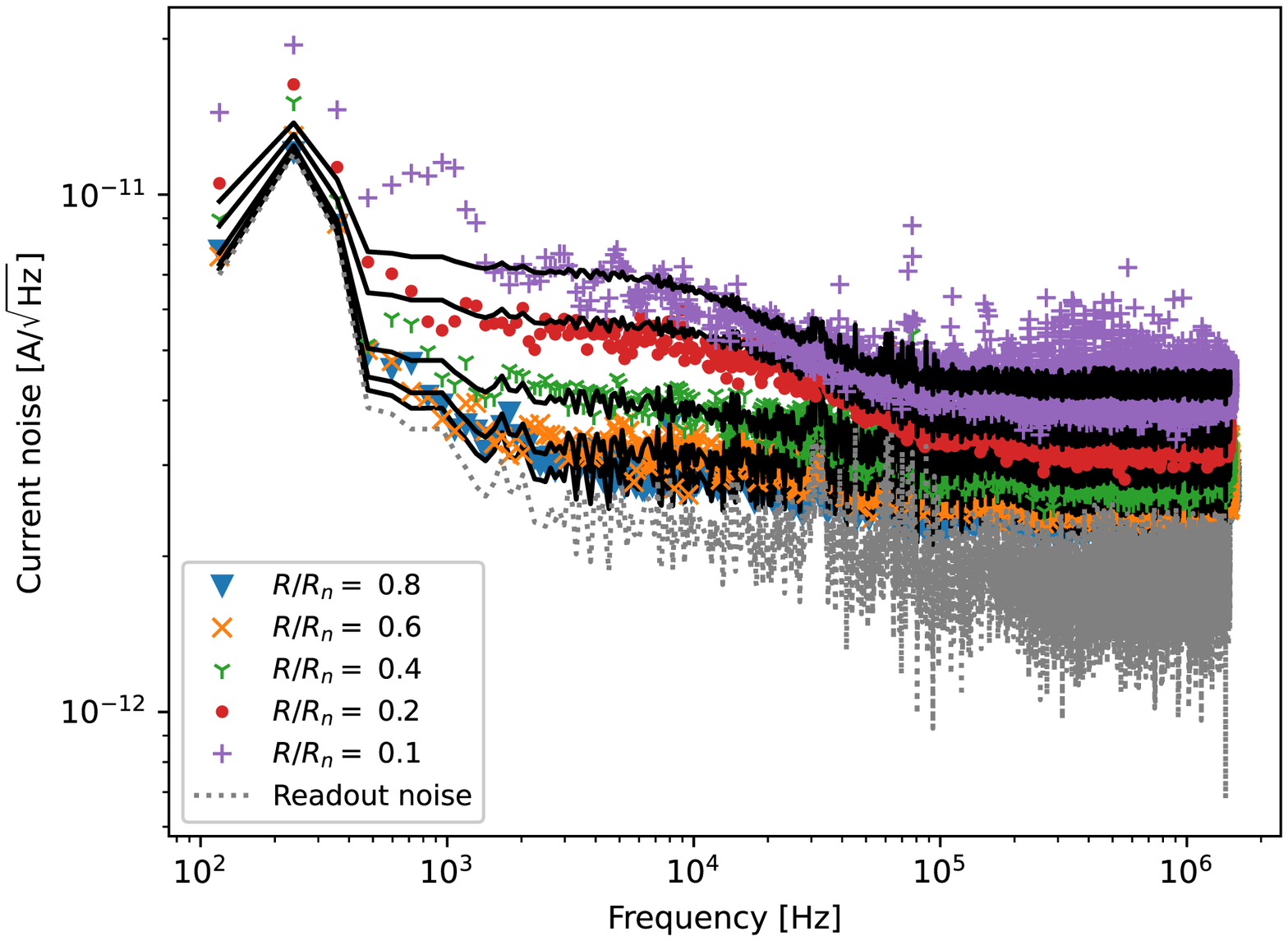}}
\hfil
\caption{ Current noise of TES1.
(a) The noise at 0.2 $R_{\rm n}$. TFN and Johnson noise were calculated using parameters extracted from the complex impedance shown in Fig.~\ref{fig:imp-TES}(a).
The temperature at the shunt resistor are assumed to be same as the bath temperature.
(b) The noise at several bias points. The black lines show the sum of the TFN, the Johnson noise, the shunt resistor Johnson noise and the readout noise. The excess noise term $M$ as a function of a bias point was taken into account.
}
%python3 plot_noise2.py noise-0p2Rn-18uA_noise_spec.dat  unit4ch1-6p7mK-20200702_2.dat results_20200713.dat results_IM_20200713.dat noise-normal-150uA_noise_spec.dat noise-0p8Rn-49uA_noise_spec.dat 49.0 noise-0p6Rn-36uA_noise_spec.dat 36.0 noise-0p4Rn-27uA_noise_spec.dat 27.0 noise-0p2Rn-18uA_noise_spec.dat 18.0 noise-0p1Rn-12uA_noise_spec.dat 12.0
%python3 plot_noise2.py noise-0p1Rn-12uA_noise_spec.dat  unit4ch1-6p7mK-20200702_2.dat results_20200713.dat results_IM_20200713.dat noise-normal-150uA_noise_spec.dat noise-0p2Rn-18uA_noise_spec.dat 18.0 noise-0p1Rn-12uA_noise_spec.dat 12.0
\label{fig:noise-TES1}
\end{figure*}

\subsection{Thermodynamic noise}
\label{sec:noise}
\begin{figure*}[!t]
\centering
\subfloat[]{\includegraphics[width=0.45\linewidth, keepaspectratio]{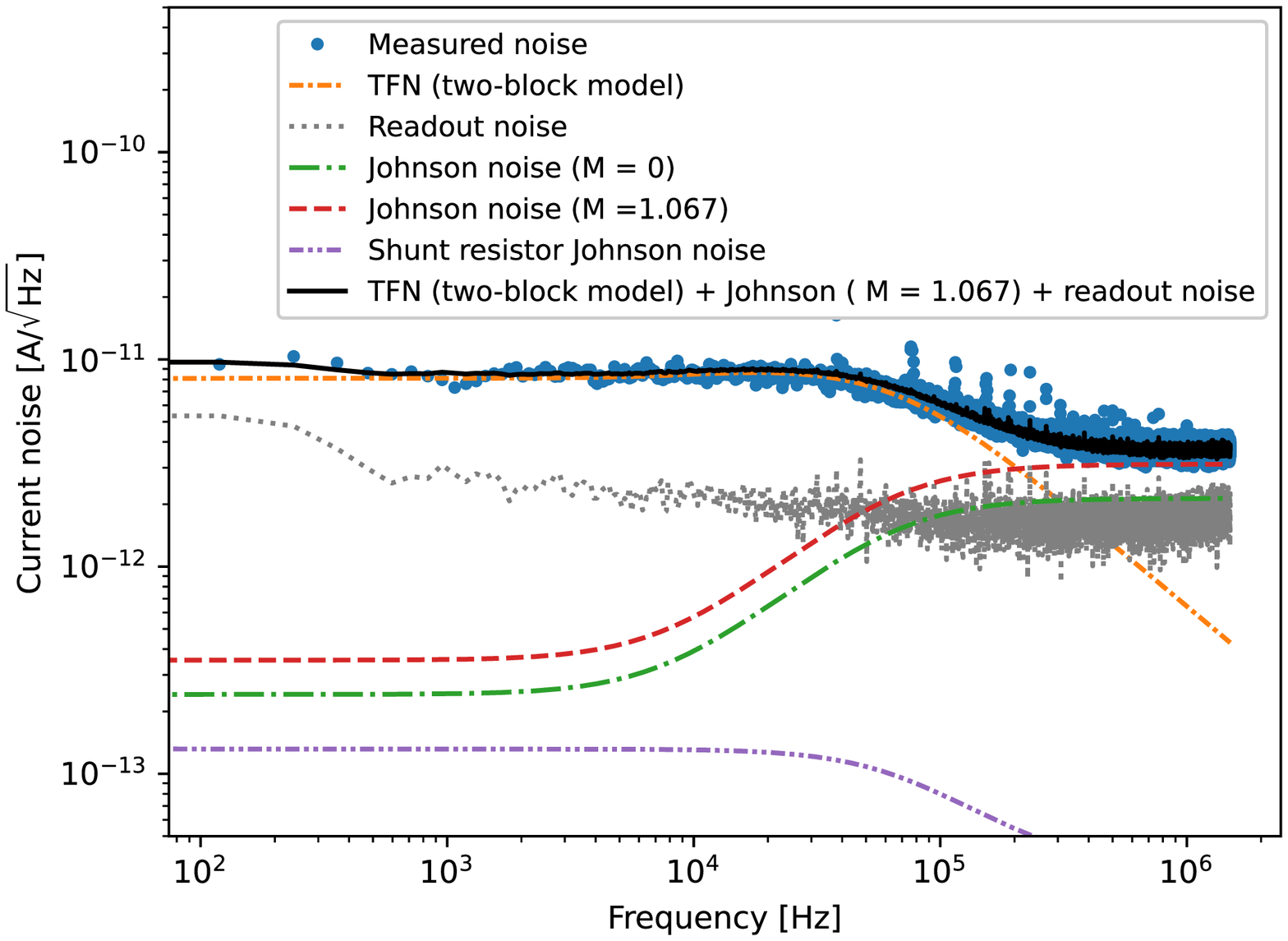}}
\hfil
\subfloat[]{\includegraphics[width=0.45\linewidth, keepaspectratio]{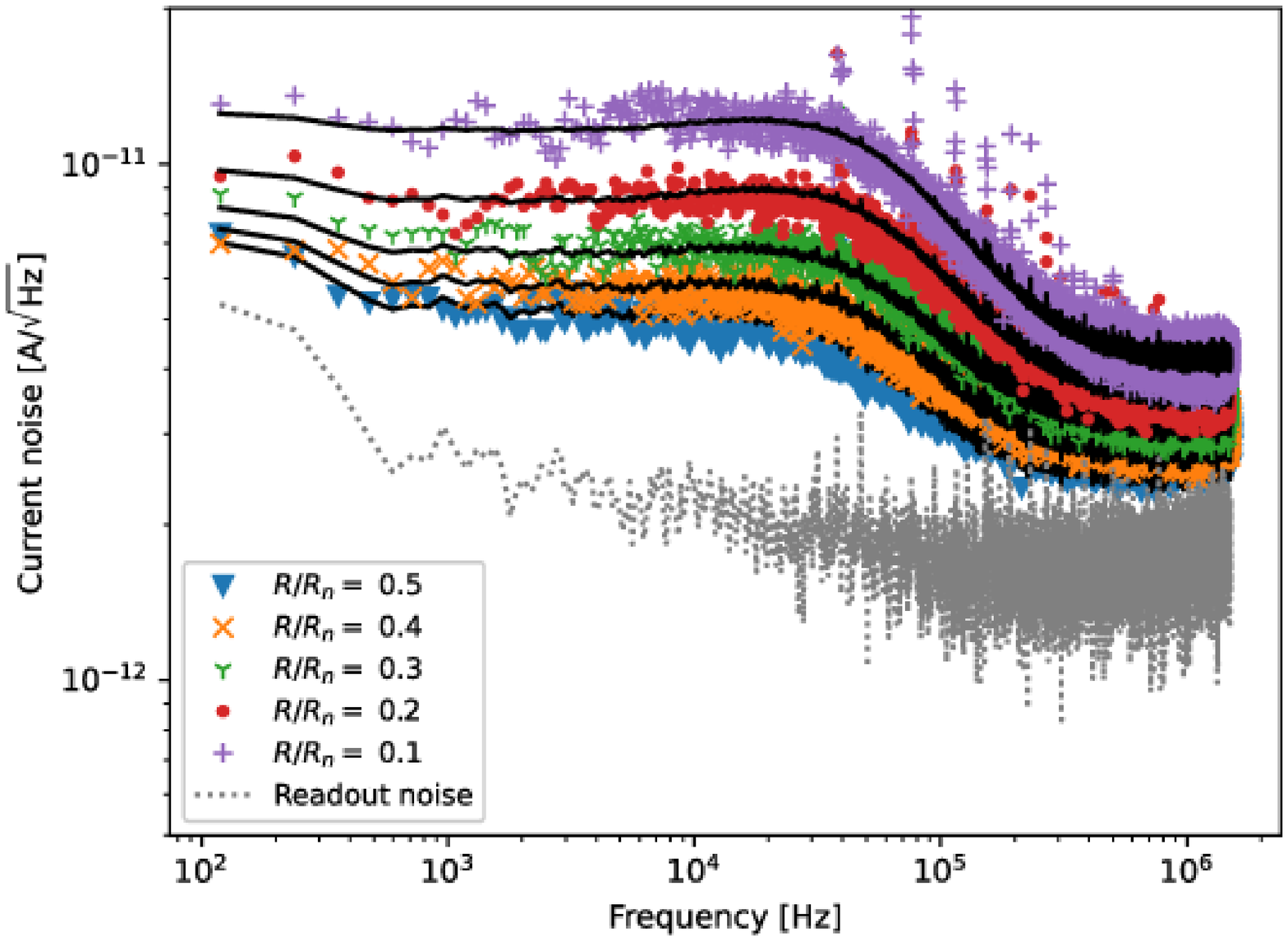}}
\hfil
\caption{ Current noise of TES2.
To calculate theoretical values of the noise, the parameters extracted the complex impedance ($\beta$, ${\cal{L}}_{\rm eff}$, $\tau_0$ and $\tau_1$), and the measured readout noise were used. 
The temperature at the shunt resistor are assumed to be same as the bath temperature.
(a) The noise at 0.2 $R_{\rm n}$.
(b) The black lines show fits to Eq.~\ref{eq:noise-two-block-model}.
}
%data/DR-Run9-20200616/WF/unit3ch1/6p7mK-20200701
%python3 plot_noise2.py IV-unit3ch1-6p7mK.dat results_20210823.dat results_IM_20200708.dat noise-normal-100uA_noise_spec.dat noise-0p5Rn-22uA_noise_spec.dat 22.0 noise-0p4Rn-20uA_noise_spec.dat 20.0 noise-0p3Rn-17uA_noise_spec.dat 17.0 noise-0p2Rn-14uA_noise_spec.dat 14.0 noise-0p1Rn-10uA_noise_spec.dat 10.0
\label{fig:noise-TES2}
\end{figure*}

In this section, we will 
%determine an appropriate model for TES2 and 
compare the measured noise with the theoretical noise level to evaluate the excess Johnson noise.
The results of noise measurements for  TES1 and TES2 are shown in Figs.\ref{fig:noise-TES1} and \ref{fig:noise-TES2}.

The theoretical noise requires the thermal conductance between a TES and the heat bath as an input parameter. The parameter is used for calculation of the thermal fluctuation noise (TFN)~\cite{ref:TES0}. The thermal conductance is associated with Joule power dissipation of the a TES. The power dissipation is equal to the power flow to the heat bath and can be written as 

\begin{equation}
P_J = K(T^n-T^n_{\rm bath}), 
\label{eq:P_J}
\end{equation}

where $n$ is the exponent determined by the nature of the thermal link to the heat bath. The parameter $K$ is associated with the thermal conductance by $G = nKT^{n-1}$. The cool-down mechanism of the TESs can be explored by measuring the Joule power dissipation as a function of the bath temperature. The measured $P_J$ for TES1 and TES2,  when they were biased at 0.5 $R_{\rm n}$, were fitted with Eq.\ref{eq:P_J} and the resultant exponents were $n = 5.07$ and $4.95$, respectively. This implies that the TESs were cooled down by electron-phonon coupling.

The measured thermal conductance of TES1 ($G_{\rm tes,b}$) was $9.9\times 10^{-12}$ W/K. 
For TES2, which agreed well with the two-block model, the measured value of $6.3\times 10^{-12}$ W/K was a function of thermal conductance in the system ($G_{\rm tes,b}$, $G_{\rm 1,b}$ and $G_{\rm tes,1}$). 
%To determine what the measured conductance represents, it was necessary to identify a model from its current noise.

In the following, we will show that the current noise in TES2 was fitted well with the two-block model. In this model, we can define the effective thermal conductance $G_{\rm eff}$, which was equal to the thermal conductance extracted from Joule power dissipation.
%,and could be equal to $G_{\rm tes,b} + G_{\rm tes,1}$ or $G_{\rm tes,b} + G_{\rm 1,b}$. 
We will show that we can calculate the theoretical current noise without knowing $G_{\rm tes,b}$ or the exact form of $G_{\rm eff}$.
%data/DR-Run7-20200328/IV/unit4ch1
%Rfrac = 0.25
%data/DR-Run9-20200616/IV/unit3ch1
%Rfrac = 0.5

In order to calculate the theoretical current noise, we also need the responsivity of the TES to small signals. The responsivity $s_I(\omega)$ can be associated with the complex impedance $Z_{\rm TES}$ by~\cite{ref:Ztes_Maasilta}

\begin{equation}
    s_I(\omega) =
    - \frac{1}{(Z_{\rm tes} + R_{\rm sh} + i\omega L)I_0}
    \frac{Z_{\rm tes} - R_0(1+\beta_I)}{R_0(2+\beta)},
    \label{eq:responsivity}
\end{equation}

where $R_sh$ is the shunt resistance in the bias circuit and 18 m$\rm \Omega$ in our setup.
The Johnson noise is written as

\begin{equation}
|I_{\omega}|^2_{\rm J} = 4kT_{\rm TES}\frac{(1+2\beta)(1+M^2)}{R_0(2+\beta)}\left|
\frac{Z_{\rm tes}+R_0}{Z_{\rm tes} +R_{\rm sh} + i\omega L}
\right|^2.
%(1+\omega^2\tau^2)%|dI/dP|^2 / {\cal L^2},
\label{eq:I_J}
\end{equation}

%where $M$ expresses the excess Johnson noise term. 
The Johnson noise due to the shunt resistor
in the bias circuit is 
$|I_{\omega}|^2_{\rm sh} = 4kT_{\rm sh}R_{\rm sh}/|Z_{\rm tes}+R_{\rm L} + i\omega L|^2$,
where $T_{\rm sh}$ is the temperature of the shunt resistor, $L$ is the SQUID input impedance and was 18 nH in our readout.
The contribution of $|I_{\omega}|^2_{\rm sh}$ to the total
current noise was negligible as shown in Figs.~\ref{fig:noise-TES1}(a) and \ref{fig:noise-TES2}(a). It should be noted that the temperature at the shunt resistor are assumed to be same as the bath temperature.

Next, we consider the TFN current noise of the single-block model.
It is written as

\begin{equation}
|I_{\omega}|^2_{TFN} = 4kT_{\rm TES}^2G_{\rm tes,b}F(T_{\rm tes}, T_{\rm bath})\left| s_I(\omega) \right|^2 
\equiv P_{\rm tes,b} \left| s_I(\omega) \right|^2 ,
\end{equation}

where $P_{\rm tes,b}$ is the power spectral density of the TFN. $F(T_{\rm tes}, T_{\rm bath})$ depends on the nature of the thermal link. In the radiative limit, it becomes $F(T_{\rm tes}, T_{\rm bath}) = \left[(T_{\rm bath}/T_{\rm tes})^{n+1}+1\right]/2$~\cite{ref:TFN-noise}.

\begin{figure*}[!t]
\centering
{\includegraphics[width=0.45\linewidth, keepaspectratio]{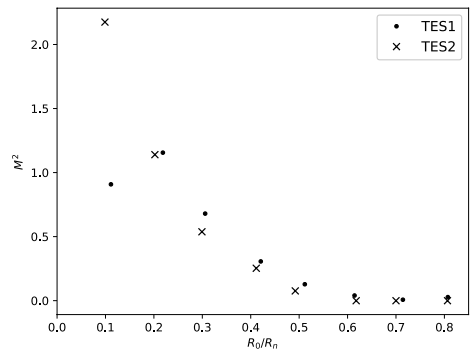}}
\caption{Excess noise parameter $M^2$ as a function of a bias point.
}
%python3 plot_data.py unit4ch1/M.dat unit3ch1/6p7mK-20200701/M.dat
\label{fig:M}
\end{figure*}

The measured current noise in TES1 shown in Fig.\ref{fig:noise-TES1}
agreed well with the sum of the theoretical noise and the readout noise. The theoretical noise was calculated using the measured thermal conductivity and parameters extracted from the complex impedance. The theoretical noise was sum of the TFN noise, the Johnson noise including the excess term $M$, and the shunt resistor Johnson noise.
%The excess Johnson noise and readout noise were also included in the noise calculation. 
$M$ was calculated at each bias point as shown in
Fig.~\ref{fig:M}.

%The two-block models give the different forms of TFN as a function of the frequency, but the same form of the complex impedance.
The TFN of the two-block model can be written as~\cite{ref:Ztes_Maasilta}

\begin{equation}
  |I_\omega|^2_{\rm TFN} = |s_I(\omega)|^2
  \left[P_{\rm tes,b}^2+P_{\rm tes,1}^2 + \frac{P_{\rm tes,1}^2}{1+\omega^2\tau_{1}^2}\left(\frac{\left(\frac{P_{\rm 1,b}G_{\rm tes,1}(T_1)}{P_{\rm tes,1}}\right)^2 + G_{\rm 1,b}^2}{(G_{\rm tes,1}(T_1)+G_{\rm 1,b})^2}-1\right)\right],
  %+ |I_\omega|^2_{J} + |I_\omega|^2_{sh},
        \label{eq:noise-two-block-model}
\end{equation}

where 
%$G_{tes,1}$ is the thermal conductance between the TES and the additional thermal block, $G_{1,b}$ is that between the additional thermal block and the thermal bath, 
$P_{tes,1}$ is the TFN between the TES and the additional thermal block, and $P_{1,b}$ is that between the additional block and the the thermal bath.
%In the parallel model, $|I_\omega|^2_{TFN}$ has $\omega^2\tau_1^2$ only in the denominator as in Eq.~\ref{eq:noise-two-block-model}. $|I_\omega|^2_{TFN}$ in other two-block models has $\omega^2\tau_1^2$ both in the denominator and the numerator~\cite{ref:Ztes_Maasilta}.
%Therefore, measured TFN as a function of the frequency can be used to determine a thermal model.

As shown in Fig.~\ref{fig:noise-TES2}(b), the measured noise was fitted well to Eq.~\ref{eq:noise-two-block-model}. 
%Therefore, the response of TES2 can be represented by the parallel model.
The parameters which can be extracted from the fit are $P_1 \equiv P_{\rm tes,b}^2+P_{\rm tes,1}^2$ and 
$P_2 \equiv 
P_{\rm tes,1}^2 \left(\frac{\left(\frac{P_{\rm 1,b}G_{\rm tes,1}(T_1)}{P_{\rm tes,1}}\right)^2 + G_{\rm 1,b}^2}{(G_{\rm tes,1}(T_1)+G_{1,b})^2}-1\right)$.
We found $P_1/P_{\rm TFN,0}^2 = 4.12$ and $P_2/P_{\rm TFN,0}^2 = -1.15$, 
where $P_{\rm TFN,0} = \sqrt{2kT^2G_{\rm eff}((T_{\rm bath}/T)^{n+1})+1)}$ is
the TFN assuming that TES2 follows the single-block model and its
thermal conductance to the heat bath is $G_{\rm eff}$.
It should be noted that $\tau_1$ extracted from the complex impedance was used for the fits.
To determine $P_{\rm tes,b}$, $P_{\rm tes,1}$, $G_{1,b}$, $T_1$ individually,
additional experiments may be necessary. 
Without knowing these values, the measured noise was able to be fitted to Eq.~\ref{eq:noise-two-block-model}, and the excess Johnson noise term was obtained as shown in Fig.~\ref{fig:M}.

%As far as reported, TES2 was the only optical TES whose response was represented by the two-block model. The complex impedance of other optical TESs was fitted well with the single-block model.%~\cite{ref:optical_TES_1MHz}~\cite{ref:Ztes-2020}. 
%TES2 may have factors that altered thermodynamics. It should be noted that a thermal model may be uncorrelated with the loop gain, which is essential for high energy resolution. We previously tested a TES whose behavior was explained by the single-block model and obtained a high loop gain comparable to TES2~\cite{ref:Ztes-2020}. Therefore, the loop gain may be independent of a thermal model.

%def model_two_body3_2(fin, ratio_ph, g):
%popt2 [-1.31894777  2.02582374]
%ratio_ph = -1.31894777, g = 2.02582374
%sqrt(abs(ratio_ph)) = 1.148454

%\subsection{Excess Johnson noise}
%We have determined the thermal models for the TESs, have showed that TFN was negligible at high frequencies and calculated the excess Johnson noise term $M^2$, as shown in Fig.~\ref{fig:M}. TES1 and TES2 have the similar trend in $M^2$ as a function of a bias point.

\subsection{Thermal models}
\begin{figure*}[!t]
\centering
\subfloat[]{\includegraphics[width=0.2\linewidth, keepaspectratio]{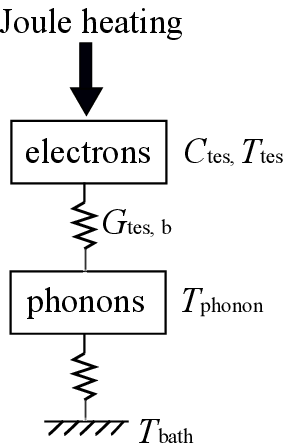}}
\hfil
\subfloat[]{\includegraphics[width=0.56\linewidth, keepaspectratio]{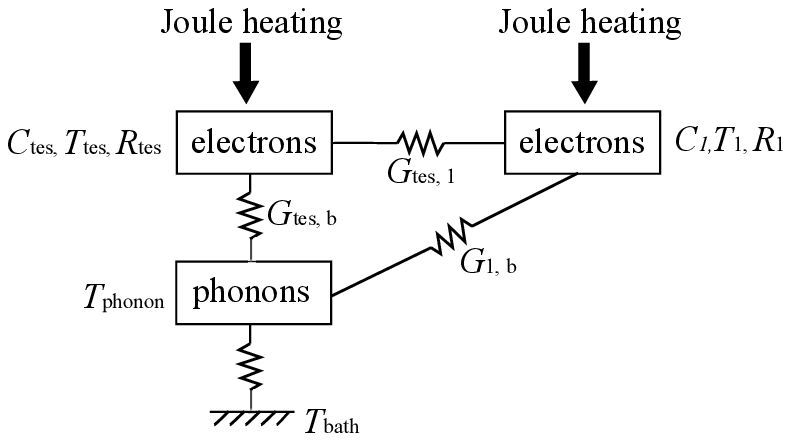}}
\hfil
\caption{ Schematic diagrams of models for (a) TES1 and (b) TES2.
%(a) A single-block model. (b) A two-block model, which is one of three two-block models.
In both TESs, the relaxation time of phonons is much faster than probe signals for measuring the complex impedance. Therefore, the phonon system can be effectively treated as the thermal bath.
}
\label{fig:thermal-model-real}
\end{figure*}

In this section, we will discuss possible models that could explain the behavior of the TESs based on detector physics. In complex-impedance measurements, response of TES1 was effectively explained by the single-block model as shown in Fig.~\ref{fig:thermal-model}(a). 
The more detail model of TES1 is presented by a thin metal film directly deposited on an insulated substrate, as shown in Fig.~\ref{fig:thermal-model-real}(a)~\cite{ref:hot-electron}. 
The relaxation time of phonons in the TES was thought to be much faster than probe signals for measuring the complex impedance. Therefore, the phonon system can be effectively treated as the heat bath, and the detector response was described well by the single-block model. 

In case of TES2, its thermal model became more complicated. From the measured complex impedance and the current noise, thermodynamic model was determined to be the two-block model. However, the components of the additional block (electrons or photons, etc) remained unknown. Here, we present a possible model that could explain the behavior of TES2.

Considering that the time constant of the additional block ($\tau_1$) was similar to that of the TES ($\tau_I$) when the loop gain was zero, it is straightforward to assume that the block could be an electron system. When the system is cooled by electron-phonon weak coupling, the time constant is determined by the temperature. Therefore, the temperature must be similar to $T_{\rm TES}$. 

To maintain the temperature above the bath temperature, the electron system must be heated. In our setup, the only mechanism that could heat the electron system was Joule dissipation. Thus, the additional block should have the resistance $R_1$. To maintain consistency with the two-block model, the current flowing through the block must be negligible and did not affect electro-thermal feedback (ETF) of the TES. This implies that $R_1$ must be much larger than $R_{\rm TES}$ and was connected in parallel with the TES. The model could have the form shown in Fig.~\ref{fig:thermal-model-real}(b). The large resistance could be created by a partial volume in the TES. The volume should be normal regardless of the voltage applying to the TES. The large resistance may imply that the cross section of the volume was small.

We will show that large resistance in parallel with a TES does not affect ETF and the set of equations are reduced to that derived by the two-block model.
When a TES is in parallel with resistance $R_1$, the electrical differential equation is

\begin{equation}
L\frac{d}{dt}(I+I_1) = V - \left(R_{L} + \left(\frac{1}{R} + \frac{1}{R_1}\right)^{-1}\right)(I+I_1),
\end{equation}

where $I_1$ are current flowing through $R_1$ and $V$ is the voltage across the bias circuit.
Here we assume that $R_1$ is constant. 
%For small signals, change in the total resistance $\Delta R_{\rm total}$ is

%\begin{equation}
%\Delta R_{\rm total} =
%\left(\frac{R_1}{R+R_1}\right)^2\Delta R,
%\end{equation}

%where $\Delta R$ is change in the resistance of the TES.
The linearized differential equation is

\begin{equation}
L\frac{d}{dt}(\Delta I + \Delta I_1) =
\Delta V - (\Delta I + \Delta I_1)R_{\rm L}
- 
%\left(\frac{R_1}{R+R_1}\right)^2 
\left(\frac{\alpha_I RI}{T_{\rm tes}}\Delta T_{\rm tes}
+ (1+\beta)R\Delta I
\right).
\label{eq:ETF-R1}
\end{equation}

when $R_1 >> R$, in frequency domain, $\Delta I_{1,\omega} = \frac{Z_{\rm tes}(\omega)}{R_1}\Delta I_{\omega}$ and thus $\Delta I_1$is much smaller than $\Delta I$.
When $|\Delta I_1| << |\Delta I|$, Eq.\ref{eq:ETF-R1} can be reduced to the equation without the resistance $R_1$.
Therefore, the contribution of $I_1$ to electro-thermal feedback is negligible. 

Next, we will show that $\Delta I_1$ does not affect the thermal response of the additional body. The linearized equations regarding to the additional body is

\begin{equation}
C_1 \frac{d\Delta T_1}{dt} = G_{\rm tes,1}(T_{\rm tes})\Delta T_{\rm tes}  - \left(G_{\rm 1,b}(T_1) + G_{\rm tes,1}(T_1)\right)\Delta T_1 + 2V_{\rm tes}\Delta I_1,
\label{eq:C_1_dt}
\end{equation}

where $V_{\rm tes}=RI=R_1I_1$ is the voltage across the TES. To show that $V_{\rm tes}\Delta I_1$ is negligible in Eq.~\ref{eq:C_1_dt}, one needs to consider the linearized equation regarding to the time derivative of the change in temperature of the TES,

\begin{equation}
C_{\rm tes} \frac{d\Delta T_{\rm tes}}{dt} 
= V_{\rm tes}(2+\beta)\Delta I + \left(\frac{\alpha RI^2}{T_{\rm tes}} - G_{\rm tes,1}(T_{\rm tes}) - G_{\rm tes,b}(T_{tes})\right)\Delta T_{\rm tes} + G_{\rm tes,1}(T_1)\Delta T_1.
\label{eq:C_tes_dt}
\end{equation}

Comparing Eqs.~\ref{eq:C_1_dt} and \ref{eq:C_tes_dt}, the terms $G_{\rm tes,1}(T_{\rm tes})\Delta T_{\rm tes} - G_{\rm tes,1}(T_1)\Delta T_1$ should be the same order of $ V_{\rm tes}(2+\beta)\Delta I$, which is much larger than $V_{\rm }\Delta I_1$. Thus, the term regarding to $\Delta I_1$ in Eq.~\ref{eq:C_1_dt} is negligible.
Therefore, the behavior of a TES in parallel with large resistance can be described by the two-block model which assumes that an additional body is not heated by current flowing through it.
The Johnson noise of $R_1$ is also negligible, which is obvious from Eq.\ref{eq:I_J}.

\subsection{Energy resolution}
We examined the measured current noise of the TESs and found that it matched with the sum of the TFN and the Johnson noise with the excess noise term implemented. Now we can readily calculate the expected energy resolution based on the current noise and can compare it to the measured energy resolution. In the small-signal limit, the FWHM energy resolution of a TES
is associated with the current noise as follows~\cite{ref:TFN-noise},

\begin{equation}
\Delta E_{\rm FWHM} = 2\sqrt{(2\ln 2)}\left[\int_0^{\infty} \frac{4}{|I_{\omega}|_{\rm total}^2}\left|s_I(\omega) \right|^2 df\right]^{-1/2}.
\label{eq:EFWHM}
\end{equation}

Using Eq.~\ref{eq:EFWHM} and the measured current noise, 
we calculated the expected energy resolution as shown in Tab.~\ref{tab:energy_resolution}.
The calculated energy resolution in the presence of the excess noise ($E_{\rm c}$) was slightly worse than the energy resolution in the absence of the excess noise.
%The calculated energy resolution in the absence of the excess noise was slightly better than the energy resolution in the presence of the excess noise.
Although the excess noise contributed significantly to the current noise, it could be excluded as a major source that limits the energy resolution.
%error in the pulse height of a signal. 
Note that the calculated energy resolution took into account the readout noise.

To compare these values with the measured energy resolution, we need to consider the energy collection efficiency. It is defined as the ratio of the energy absorbed by a TES to the energy of a photon. The collection efficiency is less than 100 \% due to phonon escape from the detector. In a linear detector, the energy deposited on a TES should be removed by electrothermal feedback and is equal to $E_{\rm ETF} = -\int_0^{\infty}V_{\rm tes}(t)\delta I(t)dt$. The energy collection efficiency becomes $E_{\rm ETF}/E_{\rm photon}$, where $E_{\rm photon}$ is the energy of a photon.
As shown in Tab.~\ref{tab:energy_resolution}, TES2 showed better efficiency than TES1 did.

If the current noise is only the source for fluctuations in the measured energy of a photon, the measured energy resolution $\Delta E_{\rm m}$ should agree with the calculated energy resolution obtained by $\Delta E_{\rm c}/\eta$. However, as in Tab.~\ref{tab:energy_resolution}, $\Delta E_{\rm c}/\eta$ is significantly smaller than $\Delta E_{\rm}$. 
The difference can be written as $\sqrt{\Delta E_{\rm m}^2 - \Delta E_{\rm c}^2/\eta^2}\equiv \Delta E_{\rm other}$. 
There must be some unexplained degradation in the energy resolution.
In TES2 whose energy resolution is less than 100 meV, $\Delta E_{\rm other}$ had a significant impact on the energy resolution..
Therefore, it is necessary to prioritize the reduction of $\Delta E_{\rm other}$ to obtain higher energy resolution..
To do this, we need to understand nature of the unexplained sources.

%We have closely speculated the current noise in the TESs. Another important noise source could be downconversion phonons~\cite{ref:phonon-escape}. Considering the phonon noise, the energy resolution is given by

%\begin{equation}
%\Delta E_{\rm FWHM} = \sqrt{E_{\rm c}^2/\eta^2 + J(E)E},
%\label{eq:EFWHM}
%\end{equation}

%where $J(E)$ is a downconversion phonon noise factor as a function of the energy of an incident photon.
%Assuming that there are only the current noise and the downconversion phonon noise, we have $\sqrt{\Delta E_{\rm m}^2 - \Delta E_{\rm c
%}^2/\eta^2} = \sqrt{J(E)E} \equiv \Delta E_{\rm ph}$. $\Delta E_{\rm ph}$ gives the lower limit on the energy resolution of a TES assuming that TFN, Johnson and readout noise are zero. 
%This implies that detector design optimization such as lowering $T_{\rm c}$ and reducing the heat capacity of a TES does not give better energy resolution than $\Delta E_{\rm ph}$.

\begin{table*}[htb]
  \caption{Energy resolution at $0.1R_{\rm n}$.}
  \centering
  \begin{tabular}{|c|c|c|} \hline
    & TES1 & TES2\\ \hline
    Measured energy resolution ($\Delta E_{\rm m}$) [meV] & 156 & 67\\ \hline
    Energy resolution calculated from the measured current noise ($\Delta E_{\rm c}$) [meV] & 89 & 40 \\ \hline
    Energy resolution without excess Johnson noise [meV] & 81 & 34\\ \hline
    Energy collection efficiency $\eta$ [\%] & 63 & 88 \\ \hline
    $\Delta E_{\rm c}/\eta$ [meV] & 141 & 46\\ \hline
    $\Delta E_{\rm other}$ [meV] & 67 & 50\\ \hline
    \end{tabular}
    \label{tab:energy_resolution}
\end{table*}

\subsection{Toward higher energy resolution}
From the noise measurements, we have found that the optical TESs have the excess Johnson noise. The effect of the excess noise on the energy resolution of TES2 was relatively small. It was strongly affected by fluctuations caused by unexplained sources. 

It had been considered that the reduction of the current noise is crucial to improve the energy resolution of an optical TES. Therefore, a TES should have low $T_{\rm c}$ and $\beta$, a small heat capacity, high $\alpha_I$ and low excess noise. We have successfully fabricated such a TES.

The study of the excess Johnson noise may be important for improvement of the energy resolution, though its contribution was small as shown in Tab.\ref{tab:energy_resolution}. For TESs designed to detect X-ray photons, the excess Johnson noise is known to be one of the limiting factors. Optical TESs have a different detector design than X-ray TES: They have no membrane or absorber and an order of magnitude smaller in size. Thus, the cause of the excess Johnson noise in optical TES could also be different. Many groups have studied the excess noise in X-ray TES, but few works have been done on the excess noise in optical TESs. This is an issue that needs to be addressed in the future.

To obtain higher energy resolution,  the source of $\Delta E_{\rm other}$ should be investigated and be reduced. 
The factors of $\Delta E_{\rm other}$ could be
(1) non-linear response, (2) error in the pulse height estimate of a signal by the optimal filtering method and (3) downconversion phonon noise~\cite{ref:phonon-escape}.

\begin{figure*}[!t]
\centering
{\includegraphics[width=0.45\linewidth, keepaspectratio]{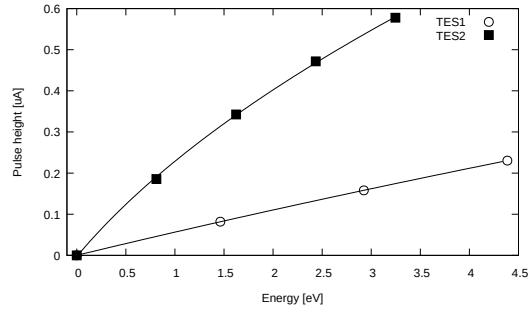}}
\caption{Pulse height as a function of the total energy of photons absorbed in a TES simultaneously. The data were fitted to a quadratic function.
}
\label{fig:pulse-height}
\end{figure*}

We discussed the energy resolution in the small-signal limit, assuming that the detector response is linear.
If there is non-linearity in a TES, the energy resolution calculated based on the small-signal theory may deviate from the measured resolution.

In a linear detector, the pulse height should be proportional to the total energy of photons.
When a linear TES is irradiated with a pulsed laser, the pulse height is proportional to the number of photons simultaneously absorbed there. 
In reality, as shown in Fig.~\ref{fig:pulse-height}, the measured pulse height was fitted to a quadratic function rather than a linear function.
TES2 showed more obvious nonlinear responses than TES1.
Note that the pulse height spectra including Fig.~\ref{fig-1550nm} were calibrated using a quadratic function, from which the energy resolution was obtained. 

%DR-Run5-20200207/WF/unit3ch1
%python3 fit_peaks_linear.py 1550nm-60dB-1GHz-tes_histogram-20211011.dat
%single-photon peak 0.06117 eV
%R-Run6-20200228/WF/unit4ch1/20200317
%python3 fit_peaks_linear.py 848nm-45dB-Ibias-10p7uA-1GHz-tes_histogram.dat.org
%single-photon peak 0.14996 eV

It is questionable whether it is optimal to fit with a quadratic function over the entire energy region. In Fig.~\ref{fig:pulse-height}, the region where TES2 responded linearly was unclear. There was a possibility that the TES responded linearly to small-energy deposition, such as a region from 0 eV to the single-photon peak. The single-photon peak in the pulse-height spectrum was fitted with Gaussian function and the resultant energy resolution was 62 meV.
The change in the resolution was 10\% from fit with the quadratic function, and implies uncertainty introduced by non-linearity. In the case of TES1, which exhibited better linearity, applying the same procedure to the measured pulse height spectrum resulted in the energy resolution of 150 meV.
The change in the resolution was 4\% and was smaller.

The non-linear responses also linked to optimal analysis of the pulse height. In the analysis, the static noise is assumed. However, a large change in the resistance (or a bias point) invalidate this assumption. This could lead to an error in the pulse height and deteriorate the energy resolution.
For future work, to understand responses of a TES in more detail,  we will study the nonlinear detector response between the small-signal limit and the large signal region where a TES is saturated. 

To improve the energy resolution, lowering $T_{\rm c}$ furthermore could be still an option in fields where slow detector response is acceptable.
In applications that require the high energy resolution, such as biological imaging and microscopic spectroscopy, the counting rate of incident photons are occasionally low, and slow detector response may be acceptable. 
For low-$T_{\rm c}$ TESs, with a given energy of a photon, a change in temperature and thus the resistance will be larger and their responses will be non-linear. Therefore, it is important to have a better understanding of the nonlinearity.

In this paper, we have focused on the current noise.
Another important source of fluctuations in the measured energy is downconversion phonon noise.
The noise is independent of the current noise and is originated from fluctuations in the number of phonons escaping into a substrate.
The noise could be reduced by forming a TES on membrane to limit escape paths of phonons.

This membrane structure was implemented by other group~\cite{ref:optical-tes-membrane}.
Its purpose was to reduce energy loss and to enhance the energy collection efficiency $\eta$. We have achieved fairly high $\eta$ without membrane. However, non-uniformity in $\eta$ was quite high, i.e., typically $\eta$ was from 50 to 90 \%. By adopting the membrane structure, we could  improve $\eta$ and the non-uniformity. 

%In general, optical TESs are formed directly on substrate to enhance detector response, which is essential for some applications, such as quantum computing and quantum information. In this field, the energy resolution should be high enough to resolve the number of photons. The energy resolution of 0.1 eV should be sufficient for this purpose. 

\section{Conclusion}
In conclusion, the energy resolution of an optical TES reached 67 meV by lowering the critical temperature $T_{\rm c}$ and by achieving a high loop gain. The measurements of complex impedance and current noise show that the behavior of the TES was consistently explained by a two-block model. We calculated the theoretical values of the current noise and compared them with the measured data, and found that there was excess Johnson noise. The excess noise term $M$ was 1.5 at a bias point where the resistance was 10\% of normal resistance. Another TES with a typical energy resolution (156 meV) was also tested and showed a similar $M$. Though the excess Johnson noise contributed to deterioration of the energy resolution, it was not a dominant source of the current noise. We found unexplained fluctuations in the measured energy. The contribution of the unexplained noise to the energy resolution was comparable with that of the current noise. 

For future work, further lowering $T_{\rm c}$ could pave the road to achieve the energy resolution below 50 meV. In parallel, we need to understand and mitigate the unexplained noise. One of possible sources of this noise could be non-linear response of the TESs. In this paper, we assumed that the small signal limit and that TESs responded linearly. The discrepancy between the measured values and the theoretical expectations could be explained by non-linearity. 
The non-linearity in detector response will become more obvious as $T_{\rm c}$ is lower and thus the heat capacity decreases. Understanding non-linearity will become more important in future research.

%\begin{acknowledgements}
\ack
This work was supported by JSPS KAKENHI Grant Number JP20K04610, JST CREST Grant Number JPMJCR2004 and JPMJCR17N4.
Transition-edge sensors were fabricated by clean room
for analog and digital superconductivity (CRAVITY) at the
National Institute of Advanced Industrial Science and Technology
(AIST). A part of this work was conducted at the AIST
Nano-Processing Facility.
%\end{acknowledgements}

\appendix
\section{Energy resolution in case of two-block model}
\label{app:E}
We will show that the energy resolution of a TES described by the two-block model is proportional to $C/\alpha_I$. 
%Here we choose the parallel model which is relevant to TES2.
%In case of other two-block models, the energy resolution can be calculated in the same manner.

Using Eqs.~\ref{eq:P_J}, \ref{eq:noise-two-block-model} and \ref{eq:EFWHM},
the energy resolution is

\begin{eqnarray}
\Delta E_{\rm FWHM} &=& 2\sqrt{(2\ln 2)}\left[\int_0^{\infty} \frac{4}
{P_1 + \frac{V^2_{\omega, tes}I_0^2(1+\omega^2\tau^2)}{{\cal{L_{\rm eff}}}}
+ \frac{P_2}{1+\omega^2 \tau^2_1}} df\right]^{-1/2} \\
&=&
2\sqrt{(2\ln 2)}
\left[\frac{1}{2\pi \tau_1 |P_2|}\int_0^{\infty} \frac{4}
{a + bx^2-\frac{1}{x^2+1}} dx\right]^{-1/2},\\
a &=& \frac{P_1}{|P_2|} + \frac{V^2_{\omega, tes}I_0^2}{|P_2|{\cal{L_{\rm eff}}}},\\
b &=& \frac{V^2_{\omega, tes}I_0^2\tau^2}{|P_2|{\cal{L_{\rm eff}}}\tau_1^2},
\label{eq:EFWHM2}
\end{eqnarray}

where $V_{\omega, tes} = \sqrt{4kT_{\rm TES}R_0(1+2\beta)}$, $I_0$ is the current flowing through the TES,
$M=0$, $x=\omega\tau$ and the strong electrothermal feedback is assumed. With the large loop gain, the complex impedance $Z_{\rm tes}$ gives the same from the single-block model. 

In the strong electrothermal feedback regime ($b << 1$), Eq.\ref{eq:EFWHM2} simplifies to

\begin{eqnarray}
\Delta E_{\rm FWHM} &=&
2\sqrt{(2\ln 2)}
\left[\frac{1}{\tau_1 |P_2|}\left[
\frac{\frac{b}{a-b}}
{\sqrt{b(1-\frac{1}{a-b})}}
+ \frac{a-b+\frac{b}{a-b}}
{\sqrt{a+\frac{b}{a-b}}}
\right] / 
(2\sqrt{b}(a-b+\frac{2b}{a-b}))\right]^{-1/2}\\
&\simeq&
2\sqrt{(2\ln 2)}
\sqrt{\tau_1 |P_2|\sqrt{ab}}\\
&=&
\sqrt{\frac{4kT^2C}{\alpha_I}\frac{P_1}{P_{\rm TFN,0}}
\sqrt{\frac{n(1+2\beta)((T_b/T)^{n+1}+1)/2}{1-(T_b/T)^n}}}.
\label{eq:EFWHM3}
\end{eqnarray}

Therefore, the energy resolution of the two-block model is proportional to $C/\alpha_I$.

\pagebreak
\bibliography{ref_TES.bib}

\begin{thebibliography}{10}

\bibitem{ref:TES0}
K.~D. Irwin and G.~C. Hilton.
\newblock Transition-edge sensors.
\newblock {\em Cryogenic Particle Detection, Topics Appl. Phys.}, 99:63--152,
  2005.

\bibitem{ref:tes-review}
J.~N. Ullom and D.~A. Bennett.
\newblock Review of superconducting transition-edge sensors for x-ray and
  gamma-ray spectroscopy.
\newblock {\em Supercond. Sci. Technol.}, 28:084003, 2015.

\bibitem{ref:optical_TES_1998}
B.~Cabrera, R.~M. Clarke, P.~Colling, A.~J. Miller, S.~Nam, and R.~W. Romani.
\newblock Detection of single infrared, optical, and ultraviolet photons using
  superconducting transition edge sensors.
\newblock {\em Appl. Phys. Lett.}, 73:735, 1998.

\bibitem{ref:quantum-computing}
E.~Knill, R.~Laflamme, and G.~J. Milburn.
\newblock A scheme for efficient quantum computation with linear optics.
\newblock {\em Nature}, 409:46--52, 2001.

\bibitem{ref:quantum-communication}
F.~E. Becerra, J.~Fan, and A.~Migdall.
\newblock Photon number resolution enables quantum receiver for realistic
  coherent optical communications.
\newblock {\em Nature Photonics}, 9:48--53, 2015.

\bibitem{ref:TES-qcommunication}
Kenji Tsujino, Daiji Fukuda, Go~Fujii, Shuichiro Inoue, Mikio Fujiwara,
  Masahiro Takeoka, and Masahide Sasaki.
\newblock Quantum receiver beyond the standard quantum limit of coherent
  optical communication.
\newblock {\em Phys. Rev. Lett.}, 106:250503, 2011.

\bibitem{ref:TES-qcommunication2}
Naoto Namekata, Yuta Takahashi, Go~Fujii, Daiji Fukuda, Sunao Kurimura, and
  Shuichiro Inoue.
\newblock Non-gaussian operation based on photon subtraction using a
  photon-number-resolving detector at a telecommunications wavelength.
\newblock {\em Nature Photon}, 4:655–660, 2010.

\bibitem{ref:quantum-metrology}
Jonathan C.~F. Matthews, Xiao-Qi Zhou, Hugo Cable, Peter~J. Shadbolt, Dylan~J.
  Saunders, Gabriel~A. Durkin, Geoff~J. Pryde, and Jeremy~L. O’Brien.
\newblock Towards practical quantum metrology with photon counting.
\newblock {\em npj Quantum Information}, 2:16023, 2016.

\bibitem{ref:quantum-metrology2}
Martin von Helversen, Jonas B{\"{o}}hm, Marco Schmidt, Manuel Gschrey,
  Jan-Hindrik Schulze, Andr{\`{e}} Strittmatter, Sven Rodt, J{\"{o}}rn~Beyer
  andTobias Heindel, and Stephan Reitzenstein.
\newblock Quantum metrology of solid-state single-photon sources using
  photon-number-resolving detectors.
\newblock {\em New J. Phys.}, 21:035007, 2019.

\bibitem{ref:tes-INRIM}
C.~Portesi, E.~Taralli, L.~Lolli, M.~Rajteri, and E.~Monticone.
\newblock Fabrication and characterization of fast tess with small area for
  single photon counting.
\newblock {\em IEEE Transactions on Applied Superconductivity}, 25:1--4, 2015.

\bibitem{ref:scanning_microscope2}
D.~Fukuda, K.~Niwa, K.~Hattori, S.~Inoue, R.~Kobayashi, and T.~Numata.
\newblock Confocal microscopy imaging with an optical transition edge sensor.
\newblock {\em J. Low Temp. Phys.}, 193:1228--1235, 2018.

\bibitem{ref:niwa-frontier}
K.~Niwa, K.~Hattori, and D.~Fukuda.
\newblock Few-photon spectral confocal microscopy for cell imaging using
  superconducting transition edge sensor.
\newblock {\em Front. Bioeng. Biotechnol.}, 9:789709, 2021.

\bibitem{ref:scanning_microscope1}
Kazuki Niwa, Takayuki Numata, Kaori Hattori, and Daiji Fukuda.
\newblock Few-photon color imaging using energy-dispersive superconducting
  transition-edge sensor spectrometry.
\newblock {\em Sci. Rep.}, 7:Art. no. 45660, 2017.

\bibitem{ref:TES-kobayashi}
Ryo Kobayashi, Kaori Hattori, Shuichiro Inoue, and Daiji Fukuda.
\newblock Development of a fast response titanium-gold bilayer optical tes with
  an optical fiber self-alignment structure.
\newblock {\em IEEE Trans. Appl. Supercon.}, 29:2101105, 2019.

\bibitem{ref:excess-noise}
N.~A. Wakeham, J.~S. Adams, S.~R. Bandler, S.~Beaumont, J.~A. Chervenak, A.~M.
  Datesman, M.~E. Eckart, F.~M. Finkbeiner, R.~Hummatov, R.~L. Kelley, C.~A.
  Kilbourne, A.~R. Miniussi, F.~S. Porter, J.~E. Sadleir, K.~Sakai, S.~J.
  Smith, and E.~J. Wassell.
\newblock Thermal fluctuation noise in mo/au superconducting transition-edge
  sensor microcalorimeters.
\newblock {\em J. Appl. Phys.}, 125:164503, 2019.

\bibitem{ref:excess-noise2}
L.~Gottardi, E.~Taralli M.~de Wit, K.~Nagayashi, and A.~Kozorezov.
\newblock Voltage fluctuations in ac biased superconducting transition-edge
  sensors.
\newblock {\em Phys. Rev. Lett.}, 126:217001, 2021.

\bibitem{ref:excess-noise3}
J.~N. Ullom, W.~B. Doriese, G.~C. Hilton, J.~A. Beall, S.~Deiker, W.~D. Duncan,
  L.~Ferreira, K.~D. Irwin, C.~D. Reintsema, and L.~R. Vale.
\newblock Characterization and reduction of unexplained noise in
  superconducting transition-edge sensors.
\newblock {\em Appl. Phys. Lett.}, 84:4206, 2004.

\bibitem{ref:optical_TES_1MHz}
E.~Taralli, L.~Lolli, E.~Monticone, M.~Rajteri, L.~Callegaro, T.~Numata, and
  D.~Fukuda.
\newblock Full characterization of optical transition-edge sensor by impedance
  spectroscopy measurements in a bandwidth extending to 1 {MH}z.
\newblock {\em ISEC 2013 Conference paper}, page DOI:
  10.1109/ISEC.2013.6604291, 2013.

\bibitem{ref:Ztes-2020}
K.~Hattori, R.~Kobayashi, S.~Takasu, and D.~Fukuda.
\newblock Complex impedance of a transition-edge sensor with sub-$\mu$ time
  constant.
\newblock {\em AIP Advances}, 10:035004, 2020.

\bibitem{ref:Lindeman_Ztes}
M.~Lindeman, S.~Bandler, R.~Brekosky, J.~Chervenak, E.~Figueroa-Feliciano,
  F.~Finkbeiner, and C.~Kilbourne M.~Li.
\newblock Impedance measurements and modeling of a transition-edge-sensor
  calorimeter.
\newblock {\em Rev. Sci. Instrum.}, 75:1283--1289, 2004.

\bibitem{ref:Ztes-akamatsu}
H.~Akamatsu, Y.~Ishisaki, A.~Hoshino, Y.~Ezoe, T.~Ohashi, Y.~Takei, N.~Y.
  Yamasaki, K.~Mitsuda, T.~Oshima, , and K.~Tanaka.
\newblock Impedance measurement of a gamma-ray tes calorimeter with a bulk sn
  absorber.
\newblock {\em AIP Conference Proceedings}, 1185:191, 2009.

\bibitem{ref:Ztes_Maasilta}
I.~J. Maasilta.
\newblock Complex impedance, responsivity and noise of transition-edge sensors:
  Analytical solutions for two- and three-block thermal models.
\newblock {\em AIP Advances}, 2:042110, 2012.

\bibitem{ref:TFN-noise}
D.~Mccammon.
\newblock Thermal equilibrium calorimeters – an introduction.
\newblock {\em Cryogenic Particle Detection, Topics Appl. Phys.}, 99:1, 2005.

\bibitem{ref:hot-electron}
F.~C. Wellstood, C.~Urbina, and John Clarke.
\newblock Hot-electron effects in metals.
\newblock {\em Phys. Rev. B}, 49:5942--5955, 1994.

\bibitem{ref:phonon-escape}
A.~G. Kozorezov, J.~K. Wigmore, D.~Martin, P.~Verhoeve, and A.~Peacock.
\newblock Resolution limitation in superconducting transition edge photon
  detectors due to downconversion phonon noise.
\newblock {\em Appl. Phys. Lett.}, 89:223510, 2006.

\bibitem{ref:optical-tes-membrane}
A.~E. Lita, A.~J. Miller, and S.~Nam.
\newblock Energy collection efficiency of tungsten transition-edge sensors in
  the near-infrared.
\newblock {\em J. Low Temp. Phys.}, 151:125--130, 2008.

\end{thebibliography}
\bibliographystyle{unsrt}

\end{document}